\documentclass[%
 reprint,
superscriptaddress,
showkeys,
nofootinbib,
amsmath,amssymb,
prf,
]{revtex4-2}

\usepackage{graphicx}% Include figure files
\usepackage{dcolumn}% Align table columns on decimal point
\usepackage{bm}% bold math
\usepackage{soul}
\newcommand{\Ca}{\text{Ca}}

\newcommand{\Bqs}{\text{Bq}_{\mbox{\scriptsize s}}}
\newcommand{\Bqd}{\text{Bq}_{\mbox{\scriptsize d}}}

\newcommand{\mus}{\mu_{\mbox{\scriptsize s}}}
\newcommand{\mud}{\mu_{\mbox{\scriptsize d}}}

\newcommand{\fmmone}{f_1^{(\mbox{\tiny MM})}}
\newcommand{\fmmtwo}{f_2^{(\mbox{\tiny MM})}}

\newcommand{\viscStress}{\boldsymbol{T}^{(\mathrm{S})}}
\newcommand{\averr}{\langle \epsilon \rangle_T}
\newcommand{\Dsteady}{D_{\mbox{s}}}

\renewcommand{\vec}[1]{\boldsymbol{\bm{#1}}} 

\begin{document}
\title{A reduced model for droplet dynamics with interfacial viscosity}
\author{F. Guglietta}
\email{fabio.guglietta@roma2.infn.it}
\affiliation{Department of Physics and INFN, Tor Vergata University of Rome, Via della Ricerca Scientifica 1, 00133 Rome, Italy}
\author{D. Taglienti}
\affiliation{Department of Physics and INFN, Tor Vergata University of Rome, Via della Ricerca Scientifica 1, 00133 Rome, Italy}
\author{M. Sbragaglia}
\affiliation{Department of Physics and INFN, Tor Vergata University of Rome, Via della Ricerca Scientifica 1, 00133 Rome, Italy}
\begin{abstract}
We propose an extension of the phenomenological Maffettone–Minale (MM) model (P.L. Maffettone and M. Minale, \textit{J. Non-Newton. Fluid Mech.} \textbf{78}, 227–241 (1998)) to describe the time-dependent deformation of a droplet with interfacial viscosity in a shear flow. The droplet, characterised by surface tension $\sigma$, is spherical at rest with radius $R$ and deforms into an ellipsoidal shape under a shear flow of rate $G$, described by a symmetric second-order morphological tensor $\boldsymbol{S}$. In addition to surface tension, the extended MM (EMM) model incorporates interfacial shear and dilatational viscosities, $\mu_s$ and $\mu_d$, through the corresponding Boussinesq numbers $\Bqs=\mu_s/\mu R$ and $\Bqd=\mu_d/\mu R$, where $\mu$ is the bulk viscosity. A central goal of this work is to quantify the parameter range over which the EMM model provides a realistic description of droplet deformation, as a function of the capillary number $\Ca=\mu R G/\sigma$ and the Boussinesq numbers. To this end, model predictions are systematically compared with fully resolved numerical simulations.
\end{abstract}
\maketitle
%%%%%%%%%%%%%%%%%%%%%%%%%%%%%%%%%%%%%%%
\section{Introduction}
\begin{figure}[t!]
\centering
\includegraphics[width=1.\linewidth]{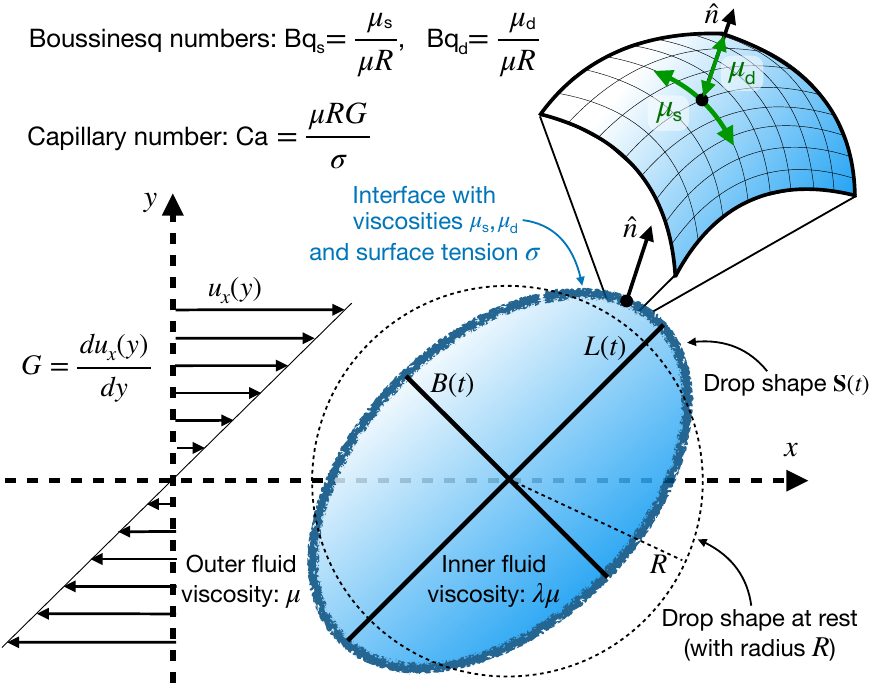}%
\caption{Shear plane section of a droplet with viscous interface deforming due to a simple shear flow with  {intensity}~$G$. The drop with radius at rest $R$, surface tension $\sigma$ and inner viscosity $\lambda\mu$ is immersed in a fluid with viscosity $\mu$. The shear interfacial viscosity $\mus$ weights the in-plane interface friction, while the dilatational interfacial viscosity $\mud$ weights the friction in the normal direction of $\vec{\hat{n}}$ (green arrows). The drop shape is described via a second-order tensor $\vec{S}(t)$, while $L(t)$ and $B(t)$ represent the major and minor axes  {in the shear plane}. } \label{fig:drop}%
\end{figure}

%%%%%%%%%%%%%%%%%%%%%%%%%%%%%%%%%%%%%%%%%%%
Understanding droplet deformation in hydrodynamic flows has been a central topic in rheology for nearly a century. 
In 1934, G. I. Taylor addressed this problem in the small-deformation limit~\cite{taylor1934formation}, analysing the response of a droplet {without interfacial viscosity} of radius $R$, surface tension $\sigma$, and viscosity ratio $\lambda$ to an external shear flow of intensity $G$ in a fluid of viscosity $\mu$ (see Fig.\ref{fig:drop}). 
Droplet deformation is characterised by the dimensionless capillary number
\begin{equation}\label{eq:Ca}
\Ca = \frac{\mu R G}{\sigma}\ ,
\end{equation}
which measures the balance between viscous stresses and surface-tension forces. Taylor’s perturbative analysis, valid for $\Ca \ll 1$, laid the foundation for subsequent theoretical developments, later refined and extended in numerous studies~\cite{Rallison1984review,Stone1984review,FischerErni2007review,cristini2004theory,Guido2011review}.
Droplet dynamics is a complex multiscale problem and in many instances it is desirable to possess a simpler phenomenological description via reduced models. 
Maffettone and Minale (MM)~\cite{maffettone1998equation} proposed an equation for the evolution of a droplet {without interfacial viscosity} in which the shape of the droplet is assumed to be ellipsoidal at all times, described via a second-order tensor $\vec{S}$. The evolution dynamics for $\vec{S}$ reads:
\begin{equation}\label{eq:MMmodel}
\begin{aligned}
\frac{d\boldsymbol{S}}{dt}
&- \left(\boldsymbol{\Omega}\!\cdot\!\boldsymbol{S}
        - \boldsymbol{S}\!\cdot\!\boldsymbol{\Omega}\right) = \\
&= - \frac{f_1}{\tau_{\sigma}}
    \left[\boldsymbol{S}-g(\boldsymbol{S})\,\boldsymbol{I}\right] 
  + f_2\left(\boldsymbol{E}\!\cdot\!\boldsymbol{S}
            + \boldsymbol{S}\!\cdot\!\boldsymbol{E}\right) \ ,
\end{aligned}
\end{equation}
where 
$\boldsymbol{\Omega}=\frac{1}{2}\left(\boldsymbol\nabla\vec{u}-\boldsymbol\nabla\vec{u}^T\right)$ and $\boldsymbol{E}= \frac{1}{2}\left(\boldsymbol\nabla\vec{u}+\boldsymbol\nabla\vec{u}^T\right)$ represent the asymmetric and symmetric velocity gradient tensors respectively, being $\vec{u}$ the fluid velocity, $\boldsymbol{I}$ the unity tensor and $g\left(\vec{S}\right)$ a nonlinear function of the $\vec{S}$ tensor invariants (see Ref.~\cite{maffettone1998equation} for details).
% Although 
Eq.~\eqref{eq:MMmodel} is formulated in a general tensorial form and can be applied to different flow configurations by specifying 
$\boldsymbol{E}$ and $\boldsymbol{\Omega}$.
In the present work, we restrict our analysis to simple shear flow.
From the tensor $\boldsymbol S(t)$, the eigenvalues associated with the length of the three principal axes of the droplet can be extracted~\cite{maffettone1998equation,Taglienti2024}. 
The deformation parameter is then computed using the two principal axes lying in the shear plane, denoted by $L(t)$ and $B(t)$, with $L>B$ (see Fig.~\ref{fig:drop}), according to
\begin{equation}\label{eq:D}
D(t)=\frac{L(t)-B(t)}{L(t)+B(t)}\ .
\end{equation}
One of the key strengths of the MM framework is that, in addition to predicting the full time evolution of the deformation, it also provides an explicit analytical expression for the stationary deformation~\cite{maffettone1998equation}. This result follows directly from the structure of Eq.~\eqref{eq:MMmodel} and is therefore independent of the specific functional forms of $f_{1,2}$, provided they are constant during the dynamics. The steady-state deformation for the simple-shear flow case is given by~\cite{maffettone1998equation}:
\begin{equation}\label{eq:Dsteady}
    \Dsteady = \frac{\sqrt{f_1^2+\Ca^2}-\sqrt{f_1^2+\Ca^2-f_2^2\Ca^2}}{f_2\Ca}\ .
\end{equation}
Concerning the coefficients $f_{1,2}$, they must be selected according to the physics of the specific system under consideration. In the original formulation, Maffettone and Minale~\cite{maffettone1998equation} focused on clean droplets {(i.e., without interfacial viscosity)}. To determine the appropriate values of $f_{1,2}$ for this case, they calibrated the model against the small-deformation framework formalised by Rallison~\cite{rallison1980note}, thus ensuring that the stationary deformation predicted by the MM equation reproduces the classical Taylor result~\cite{taylor1934formation}.  {The model parameters $f_{1,2}^{(\mbox{\tiny MM})}(\lambda)$ depend only on the viscosity ratio $\lambda$ and are given by}:
\begin{equation}\label{f1f2mm}
\small 
\fmmone(\lambda)=\frac{40(\lambda+1)}{(2\lambda+3)(19\lambda+16)},\quad 
\fmmtwo(\lambda)=\frac{5}{2\lambda+3}.
\end{equation}
Since its introduction, the MM model has become a cornerstone in the theoretical description of droplet deformation in flow, largely thanks to its ability to reproduce experimental observations across a wide range of conditions~\cite{maffettone1998equation,minale2008phenomenological,minale2010microconfined,Guido2011review,biferale2014deformation,ray2018droplets}. Over the years, several extensions have broadened its scope, including: formulations for the complex rheology of non-Newtonian fluids~\cite{minale2004deformation}, adaptations for confined geometries~\cite{minale2008phenomenological,minale2010microconfined}, variants capturing tank-treading dynamics in soft suspensions~\cite{art:arora04, scarpolini25}, generalizations valid at finite capillary numbers~\cite{taglienti2023reduced}. The MM framework has also been employed to interpret and validate fully resolved numerical simulations of droplets in homogeneous and isotropic turbulence~\cite{Taglienti2024,guglietta2025Deformation}. An extensive overview of these and other reduced approaches for droplet dynamics is available in Ref.~\cite{minale2010models}.
%%%%%%%%%%%%%%%%%%%%%%%%%%%%%%%%%%%%%%%%%%%%%%%%%%

\begin{figure}[t!]
\centering
\includegraphics[width=.95\linewidth]{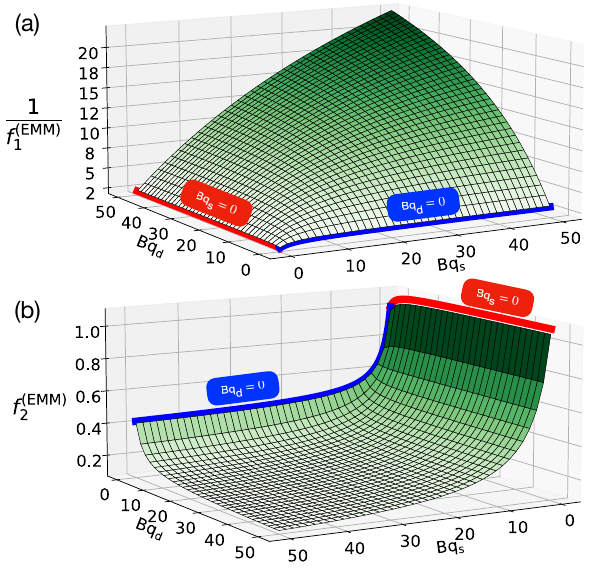}
\caption{ {EMM model parameters $1/f_1^{(\mathrm{EMM})}$ (a) and $f_2^{(\mathrm{EMM})}$ (b) as functions of $\Bqs$ and $\Bqd$ (see Eq.~\eqref{ourf1f2}).} The thick red and blue curves highlight the two boundary sections of the surface corresponding to $\Bqs=0$ and $\Bqd=0$, respectively.}\label{fig:f1}%
\end{figure}

\begin{figure}[t!]
\centering
\includegraphics[width=.95\linewidth]{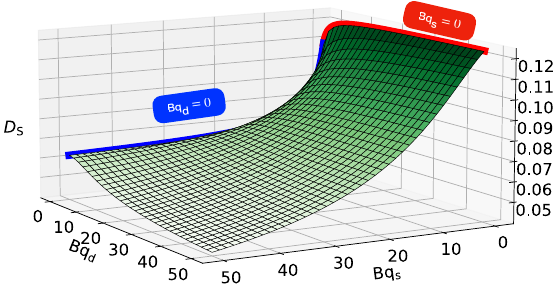}
\caption{Stationary deformation $\Dsteady$ predicted by Eq.~\eqref{eq:Dsteady} as a function of $\Bqs$ and $\Bqd$, evaluated with $f_1=f_1^{(\mathrm{EMM})}$ and $f_2=f_2^{(\mathrm{EMM})}$ as given in Eq.\eqref{ourf1f2}, for $\Ca=0.1$ and $\lambda=1$.}\label{fig:Dsteady}%
\end{figure}

Beyond surface tension effects, interfacial dynamics is often enriched by further complexities~\cite{fuller2012complex} that modify the balance of stresses between the bulk fluids. Among these, adsorption and desorption of surfactants can induce a dynamic surface tension and generate surface layers with non-negligible interfacial viscosity~\cite{sagis2011dynamic}. Interfacial viscosity introduces extra friction between the interface and the surrounding fluids, strongly influencing the deformation and relaxation of soft particles such as vesicles, capsules, and red blood cells~\cite{evans19891,hochmuth1979red,li2021similar,guglietta2020effects,guglietta2020lattice,guglietta2021loading}. Its role has also been demonstrated in drop breakup and capillary instabilities~\cite{PonceTorres2017surfact,luo2019influence,wee2020effects,herrada2021stability,graessel2021rayleigh,bacher2021rayleigh}, film stability~\cite{shen2018capillary}, and coalescence phenomena~\cite{ozan2019role,narsimhan2019shape}. Evidence of the effects of viscous interfaces motivates the inclusion of interfacial viscosity within reduced-order models of droplet dynamics. In this  {letter}, we pursue this goal by a extending the MM model to account for viscous interfaces. For a clean droplet, the dimensionless surface stress tensor, $\boldsymbol{T}^{(\mathrm{S})}$, regulating  the force balance across the
droplet interface, receives contribution only from the surface tension, $\boldsymbol{T}^{(\mathrm{S})} = (1/\Ca) \boldsymbol{P}$, where  {$\boldsymbol{P} = \boldsymbol{I} - \vec{\hat{n}}\vec{\hat{n}}$} is the tangential projection operator built from the unit normal $\vec{\hat{n}}$ to the interface~\cite{flumerfelt1980effects,narsimhan2019shape}. This stress term acts to minimize the local curvature of the interface. When the interface is viscous, an additional tangential stress appears, accounting for the resistance to shear and dilatational motion. 
{We account for both \textit{shear} and \textit{dilatational} interfacial viscosities ($\mus$ and $\mud$, respectively): $\mus$ quantifies viscous dissipation associated with tangential (shear) surface flows, while $\mud$ quantifies dissipation associated with area changes (surface dilatation and compression). They become important when the interface is coated by surfactants, polymers, or proteins, in which case surface viscous stresses can compete with capillarity and significantly affect droplet deformation dynamics and breakup thresholds~\cite{PonceTorres2017surfact,sagis2011dynamic,luo2019influence,wee2020effects,herrada2021stability,graessel2021rayleigh,bacher2021rayleigh}.}
{The effects of both interfacial viscosities} are introduced through the dimensionless Boussinesq numbers. Nondimensionalizing them with $\mu R$ ensures a consistent comparison between surface and bulk dissipative effects\footnote{Unlike bulk viscosity, interfacial viscosities have dimensions of momentum per unit length (Pa\ m\ s).}:
\begin{equation}\label{eq:bqs_bqd}
    \Bqs=\frac{\mus}{\mu R}\ ,\qquad \Bqd=\frac{\mud}{\mu R}\ .
\end{equation}
For a Newtonian interface, the extra stress contribution due to the viscous interface is provided by the Boussinesq–Scriven law~\cite{Scriven1960,langevin2014rheology}:
\begin{equation}\label{eq:stress}
\viscStress
=
\left[\frac{1}{\Ca} + (\Bqd - \Bqs)\big(\boldsymbol{\nabla}^{(\mathrm{S})}\cdot\vec{u}^{(\mathrm{S})}\big)\right]\boldsymbol{P}
+ 2\Bqs\boldsymbol{D}^{(\mathrm{S})},
\end{equation}
where $\boldsymbol{\nabla}^{(\mathrm{S})}=\boldsymbol{P}\cdot\boldsymbol{\nabla}$ denotes the surface gradient operator, and $\vec{u}^{(\mathrm{S})}$ the surface velocity field.  The symmetric surface rate–of–deformation tensor reads
$\boldsymbol{D}^{(\mathrm{S})}
= \tfrac{1}{2}
\left[
\boldsymbol{P}\cdot
\big(
\boldsymbol{\nabla}^{(\mathrm{S})}\vec{u}^{(\mathrm{S})}
+ (\boldsymbol{\nabla}^{(\mathrm{S})}\vec{u}^{(\mathrm{S})})^{\dagger}
\big)
\cdot\boldsymbol{P}
\right]$ 
and characterizes the in-plane surface deformation rate.
Such a stress tensor has been incorporated in previous analytical studies of droplet-shape evolution based on perturbative approaches~\cite{barthesbiesel1985,flumerfelt1980effects,narsimhan2019shape}. In particular, Barthès-Biesel and Sgaier~\cite{barthesbiesel1985} considered the specific case $\Bqs = \Bqd$, while Narsimhan~\cite{narsimhan2019shape} extended the analysis to the more general condition $\Bqs \ne \Bqd$. 
The latter work can be regarded as the natural counterpart of Rallison’s calculation~\cite{rallison1980note}, in which the same perturbative framework was applied to a clean interface. In direct analogy with the original MM model where Rallison’s results were used to determine the coefficients $f_{1,2}^{(\mbox{\tiny MM})}(\lambda)$ for a clean droplet (see Eq.~\eqref{f1f2mm}), we consider the calculations of Narsimhan~\cite{narsimhan2019shape}\footnote{With reference to Appendix D in~\cite{narsimhan2019shape}, $f_1=-a_D$ and $f_2=a_E$.} to derive generalized expressions for the model coefficients. These expressions constitute the natural extension of Eq.~\eqref{f1f2mm} to the case of droplets with interfacial viscosity and lead to the extended forms:
\begin{equation}\label{ourf1f2}
\small
\begin{split}
&f_1^{(\mbox{\tiny EMM})}(\lambda,\Bqs,\Bqd) =\frac{40(\lambda+1)+16(2\Bqs+3\Bqd)}{\tilde{f}(\lambda,\Bqs,\Bqd)}\ , \\
&f_2^{(\mbox{\tiny EMM})}(\lambda,\Bqs,\Bqd) =\frac{5[8\Bqs+24\Bqd+(19\lambda+16)]}{\tilde{f}(\lambda,\Bqs,\Bqd)}\ ,
\end{split}
\end{equation}
with the common denominator given by:
\begin{equation}
\begin{aligned}
\tilde{f}(\lambda,\Bqs,&\Bqd) = 32\Bqs\Bqd+4\Bqs(13\lambda+12)+\\
+2&\Bqd(23\lambda+32)+(2\lambda+3)(19\lambda+16)\ .
    \end{aligned}
\end{equation}
MM model coefficients for a clean droplet  $f_{1,2}^{(\mbox{\tiny MM})}(\lambda)$ are readily retrieved by imposing $\Bqs=\Bqd=0$ in Eq.~\eqref{ourf1f2}. Values of $1/f_1^{(\mbox{\tiny EMM})}$~\footnote{Note that, in Eq.~\eqref{eq:MMmodel}, $\tau_\sigma/f_1$ plays the role of a characteristic deformation time~\cite{maffettone1998equation}.} and $f_2^{(\mbox{\tiny EMM})}$ as functions of $\Bqs$ and $\Bqd$ are reported in Fig.~\ref{fig:f1}.
The coefficients $f_{1,2}^{(\mathrm{EMM})}$ can therefore be directly inserted into Eq.~\eqref{eq:Dsteady} to compute the stationary deformation $\Dsteady$. In Fig.~\ref{fig:Dsteady} we report a representative plot of $\Dsteady$ as a function of $\Bqs$ and $\Bqd$ for $\Ca=0.1$ and $\lambda=1$.
%------------------------------------------------------------------
%------------------
%------------------------------------------------------------------

%------------------------------------------------------------------
\begin{figure*}[t!]
\centering
\includegraphics[width=1.\linewidth]{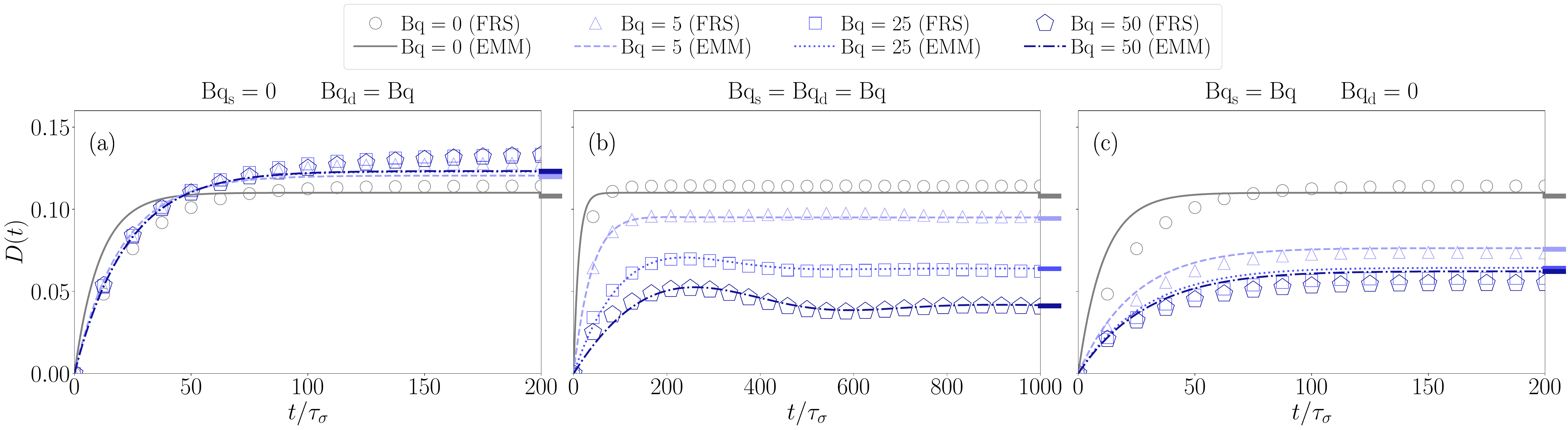}%
\caption{{Time evolution of the deformation $D(t)$ for a drop in shear flow (see Eq.~\eqref{eq:D} and  Fig.~\ref{fig:drop}) at $\Ca=0.1$ and $\lambda=1$, for different interfacial viscosities expressed by the Boussinesq numbers (see Eq.~\eqref{eq:bqs_bqd}). Time is made dimensionless using the characteristic droplet time $\tau_\sigma=\mu R/\sigma$. Fully resolved simulations (FRS) results are shown with symbols, while predictions from the numerical integration of  {Eq.~\eqref{eq:MMmodel} with $f_1$ and $f_2$ given by Eq.~\eqref{ourf1f2}} (EMM) are reported as lines. Outer horizontal ticks mark the steady deformation $\Dsteady$ from Eq.~\eqref{eq:Dsteady}. (a) $\Bqs=0$, $\Bqd=0,5,25,50$; (b) $\Bqs=\Bqd=0,5,25,50$; (c) $\Bqd=0$, $\Bqs=0,5,25,50$. {Note that (b) is shown over a longer time interval than (a) and (c) to capture the slower transient in that case.}}
} \label{fig:D_vs_t}%
\end{figure*}
%------------------------------------------------------------------
The EMM model depends on four main parameters: the viscosity ratio $\lambda$, the capillary number $\Ca$, and the two interfacial viscosity coefficients $\Bqs$ and $\Bqd$. Since the parameter space is already large and our focus is on the roles of $\Ca$, $\Bqs$, and $\Bqd$, we fix $\lambda=1$. 
Within this setting, a key question is how far the EMM model remains quantitatively accurate beyond its nominal perturbative regime.  Indeed, the EMM model is formulated under the assumption of small deformations, i.e., $\Ca \ll 1$, while the parameters $\lambda$, $\Bqs$, and $\Bqd$ are typically of order unity~\cite{narsimhan2019shape}. 
To assess this question, one can compare the predictions of the EMM model against  {reference data obtained from numerical simulations that do not rely on the small-deformation assumption.} 
Our aim is to delineate the range of $\Ca$, $\Bqs$, and $\Bqd$ over which the EMM model provides a reliable description of droplet deformation dynamics.
%%%%%%%%%%%%%%%%%%%%%%%%%%%%%%%%%%%%%%%%%%
\section{Numerical methods}\label{sec:methods}
%%%%%%%%%%%%%%%%%%%%%%%%%%%%%%%%%%%%%%%%%%
To obtain  {reference} data for droplet deformation, we perform fully resolved numerical simulations (FRS) using the immersed boundary–lattice Boltzmann method~\cite{book:kruger}. This hybrid numerical technique enables a detailed description of the coupled dynamics between the inner and outer fluids and the deformable droplet interface. The fluid motion is solved through the lattice Boltzmann method~\cite{book:kruger,benzi1992lattice}, while the interface is represented by a triangulated surface mesh that interacts with the surrounding fluid through the immersed boundary forcing~\cite{book:kruger,peskin2002immersed}. The numerical setup and implementation closely follow the procedure described in Ref.~\cite{guglietta2020effects} {(Sec. 3.1)}, which has been validated across  {a broad range of setups}~\cite{taglienti2023reduced,pelusi2023sharp,guglietta2024analytical,bellantoni2025immersed,guglietta2020effects,guglietta2021loading,guglietta2020lattice,Guglietta2023,guglietta2025Deformation}.
{For all the simulations, the fluid domain is a cubic box of size $L_x=L_y=L_z=128~\mbox{LU}$ (lattice units). A droplet of radius $R=20~\mbox{LU}$ is initialized at the box centre and its interface is discretised with a triangular mesh made of $20480$ triangles. Periodic boundary conditions are applied along the x and z directions, while along y we impose no-slip walls via bounce-back~\cite{book:kruger}. The shear flow is generated consistently with these boundary conditions.}
In the EMM model, on the other hand, the droplet shape is represented by a symmetric second-order tensor $\boldsymbol{S}$, whose evolution is governed by Eq.~\eqref{eq:MMmodel}, with the inclusion of the interfacial viscosity effects through the coefficients $f_{1,2}^{(\mbox{\tiny EMM})}$ (see Eq.~\eqref{ourf1f2}). Unlike the FRS, the EMM model does not require solving the full hydrodynamic field and is instead integrated directly from the local flow gradients using standard explicit Runge-Kutta method~\cite{taglienti2023reduced}. 
{In both cases, we measure the two main axes of the drop shape in the shear plane, namely $L(t)$ and $B(t)$ (see Fig.~\ref{fig:drop}), by retrieving them from the eigenvalues of the inertia tensor. In the EMM model, this tensor coincides with $\mathbf{S}$ (see Eq.~\eqref{eq:MMmodel}); in the FRS, the inertia tensor is computed directly from the instantaneous coordinates of the triangular-mesh nodes discretising the interface, and its eigenvalues provide $L(t)$ and $B(t)$~\cite{taglienti2023reduced,guglietta2020effects}. We then compute the deformation $D(t)$ via Eq.~\eqref{eq:D}.}
To quantify the discrepancy between the deformation obtained from FRS ($D_{\text{FRS}}(t)$) and that predicted by the EMM ($D_{\text{EMM}}(t)$), we introduce the mean relative error:
\begin{equation}\label{eq:averr}
    \averr = \frac{1}{T}\int_T\frac{|D_{\text{FRS}}(t)-D_{\text{EMM}}(t)|}{D_{\text{EMM}}(t)}dt\ ,
\end{equation}
which provides a single, global measure of the model’s accuracy. The averaging interval $T$ is selected to include both the transient and the stationary regimes of the deformation dynamics, so that the computation of $\averr$ weights the two phases in a balanced and comparable manner.
%%%%%%%%%%%%%%%%%%%%%%%%%%%%%%%%%%%%%%%%%%%%%
\section{Results and discussions}\label{sec:results}
%%%%%%%%%%%%%%%%%%%%%%%%%%%%%%%%%%%%%%%%%%%%%
We analyze the time-dependent deformation extracted from the model given in Eq.~\eqref{eq:MMmodel} with $f_{1,2}$ given in Eq.~\eqref{ourf1f2} and we compare results with FRS. 
The Boussinesq numbers affect the time-dependent droplet deformation causing deviations from the clean droplet case. This opens up for a wide variety of deformation scenarios depending on the values of $\Ca$, $\Bqs$ and $\Bqd$ (we remind that the viscosity ratio is fixed, $\lambda=1$). 
%%%%%%%%%%%%%%%%%%%%%%%%%%%%%%%%%%%%%%
\begin{figure*}[t!]
\centering
\includegraphics[width=1.\linewidth]{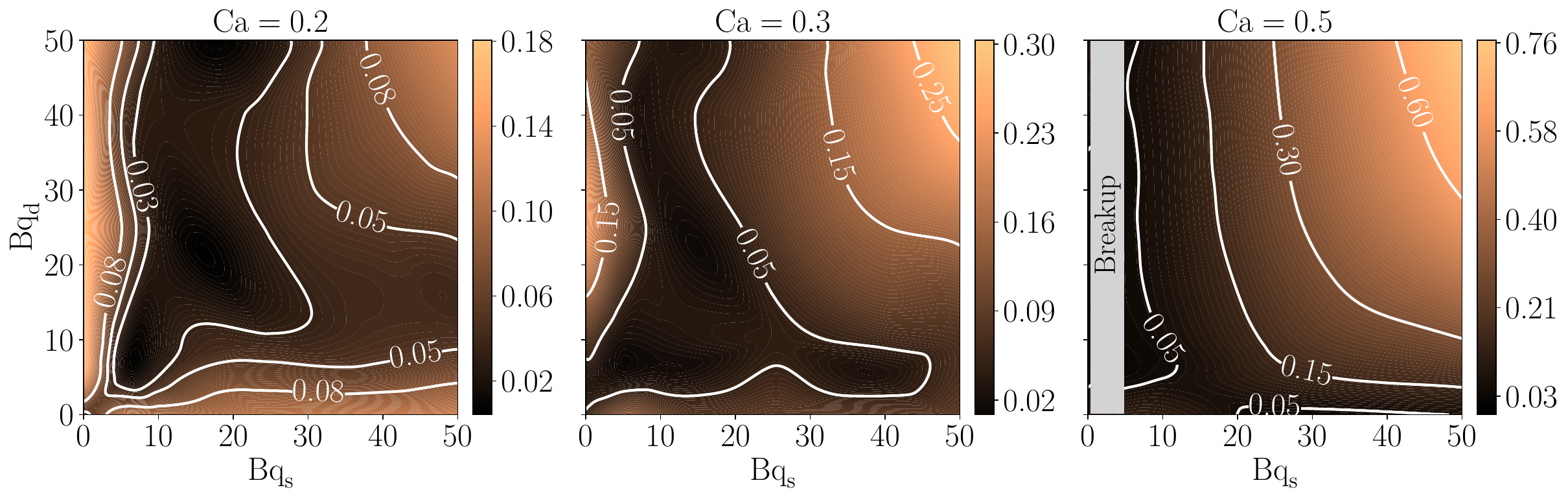}
\caption{Mean relative error $\averr$ (see Eq.~\eqref{eq:averr}) between FRS and EMM predictions for the deformation $D(t)$, shown across the ($\Bqs,\Bqd$)-plane for different capillary numbers $\Ca$.}\label{fig:error}
\end{figure*}
%%%%%%%%%%%%%%%%%%%%%%%%%%%%%%%%%%%%%%
In Fig.~\ref{fig:D_vs_t}, we report results for the time evolution of the deformation parameter $D(t)$ (see Eq.~\eqref{eq:D}) for $\Ca=0.1$. Time is made dimensionless on the x-axis via the characteristic droplet time $\tau_\sigma=\mu R/\sigma$. Similarly to previous works~\cite{art:gounley16,narsimhan2019shape,guglietta2020effects,art:lizhang19}, we analyzed three representative cases: droplet with zero shear interfacial viscosity and non-zero dilatational interfacial viscosity ($\Bqs=0$ and $\Bqd>0$, panel (a)); droplet with both interfacial viscosities of equal magnitude ($\Bqs=\Bqd>0$, panel (b)); droplet with zero dilatational interfacial viscosity and non-zero shear interfacial viscosity ($\Bqd=0$ and $\Bqs>0$, panel (c)). Data for a clean droplet ($\Bqs=\Bqd=0$) are also reported for all cases (gray lines and gray markers). 
Panels (b–c) show that increasing the interfacial viscosity leads to a reduction in the steady-state deformation, whereas panel~(a) displays the opposite trend, with a slight increase of deformation as viscosity grows. Both behaviours are consistent with previous observations~\cite{art:gounley16,guglietta2020effects,art:lizhang19,art:yazdanibagchi13}, which reported similar trends depending on the relative magnitude of the shear and dilatational interfacial viscosities. This difference originates from the structure of the viscous stress tensor $\viscStress$ (see Eq.~\eqref{eq:stress}). When $\Bqs=0$, only the surface-divergence term $\boldsymbol{\nabla}^{(\mathrm{S})}\cdot\vec{u}^{(\mathrm{S})}$ contribution remains active, enhancing the deformation~\cite{art:gounley16}. When $\Bqs=\Bqd$, such a term vanishes, leading to a monotonic decrease in deformation with viscosity.  Finally, when $\Bqd=0$, the surface-divergence term couples with the rate-of-deformation tensor $\bm{D}^{(\mathrm{S})}$, producing a behaviour that reflects the combined influence of the two terms. The EMM produces a steady-state deformation $\Dsteady$ that matches very well the FRS results and the values predicted by Eq.~\eqref{eq:Dsteady} (horizontal ticks outside the plotting area). Additional comments on the behaviour of the steady-state deformation given in Eq.~\eqref{eq:Dsteady} for different choices of the Boussinesq numbers are in order. We first note that, if $\Ca/f_1 \ll 1$ and $\Ca f_2/f_1 \ll 1$, Eq.~\eqref{eq:Dsteady} can be Taylor expanded, yielding the linear approximation $\Dsteady = \Ca f_2/(2f_1)$~\cite{maffettone1998equation}.
With reference to Fig.~\ref{fig:D_vs_t}, panels (a) and (c), we observe that in both cases  {$f_{1,2}^{(\mbox{\tiny EMM})}\sim \mathcal{O}(1)$ (see Fig.~\ref{fig:f1}), so that Eq.~\eqref{eq:Dsteady} is well represented by the linear approximation. More in details, $1/f_1^{(\mbox{\tiny EMM})}$ assumes comparable values in the two configurations (approximately $1/f_1^{(\mbox{\tiny EMM})}\simeq 2.7$), while $f_2^{(\mbox{\tiny EMM})}$ differs significantly: for $\Bqs=0$ one finds $f_2^{(\mbox{\tiny EMM})}\simeq 1$, whereas for $\Bqd=0$, $f_2^{(\mbox{\tiny EMM})}\simeq 0.4$. As a consequence, the ratio $f_2^{(\mbox{\tiny EMM})}/f_1^{(\mbox{\tiny EMM})}$ is smaller in the case $\Bqd=0$, which directly explains why the steady-state deformation $\Dsteady$ is reduced when $\Bqd=0$ compared to the case $\Bqs=0$ (see Fig.~\ref{fig:f1}, panel (b)). By contrast, in Fig.~\ref{fig:D_vs_t}, panel (b), (i.e., $\Bqs=\Bqd$) the condition $\Ca/f_1^{(\mbox{\tiny EMM})} \ll 1$ is no longer satisfied (see Fig.\ref{fig:f1}, panel (a)). As a consequence, the deformation can no longer be obtained by the linear approximation and the full nonlinear expression in Eq.~\eqref{eq:Dsteady} must be retained. 
{In summary, the range of $\Ca$ over which the linear response theory (i.e., linearity in $\Ca$) remains valid depends on the values of $\Bqs$ and $\Bqd$: in the equiviscous case ($\Bqs=\Bqd$), this range is significantly narrower than in strongly asymmetric viscosity configurations, where either $\Bqs$ or $\Bqd$ dominates.

Beyond the steady-state deformation, the EMM model reproduces well also the transient  {dynamics}. Remarkably, for $\Bqs=\Bqd$ (Fig.~\ref{fig:D_vs_t}, panel (b)), the EMM captures not only the overall increase in the characteristic deformation time, but also the transition from a monotonic relaxation to an oscillatory response (consistent with the observations in Ref.~\cite{narsimhan2019shape}) whose frequency increases with the interfacial viscosity. Moreover, 
results in Fig.~\ref{fig:D_vs_t} indicate that interfacial viscosity produces larger deformation times when $\Bqs=\Bqd$, while in the other cases ($\Bqs=0$ or $\Bqd=0$) interfacial viscosity does not alter significanlty the otherwise clean droplet deformation time. This can be understood by looking once again at Fig.~\ref{fig:f1}, panel (a), that shows the quantity $1/f_1^{(\mbox{\tiny EMM})}$, which sets the intrinsic deformation time of the droplet within the EMM framework~\cite{maffettone1998equation}. According to Eq.~\eqref{eq:MMmodel}, larger values of $1/f_1$ correspond to longer transient dynamics and slower approach to the stationary deformation. Fig.~\ref{fig:f1}, panel (a), reveals that when either $\Bqs$ or $\Bqd$ is small, the deformation time remains close to the clean-droplet value. Only when both interfacial viscosities become appreciable,  {$1/f_1^{(\mbox{\tiny EMM})}$} increases sharply. This behaviour originates from the nonlinear coupling term in the denominator of Eq.~\eqref{ourf1f2}: the relaxation rate deviates significantly from its clean-droplet value only when shear and dilatational viscosities act simultaneously. A droplet with predominantly shear viscosity or predominantly dilatational viscosity behaves almost like a clean droplet in terms of deformation time, whereas a droplet with comparable values of both viscosities experiences a dramatic slow-down.

Finally, we aim to determine the region of the $(\Ca, \Bqs, \Bqd)$ parameter space in which the EMM model remains quantitatively reliable. We performed a systematic error analysis based on the deformation evolution $D(t)$ obtained from FRS and that predicted by the EMM model. For every FRS dataset, the instantaneous relative error was evaluated and then averaged over time to obtain the mean relative error $\averr$ (see Eq.~\eqref{eq:averr}). The exploration of the Boussinesq-number space was carried out comprehensively: for every value of $\Ca$, we performed roughly 36 simulations covering a broad set of $(\Bqs,\Bqd)$ combinations. This extends previous studies where the analysis was restricted to the special cases $(\Bqs=\Bqd$, $\Bqd=0$, or $\Bqs=0$)~\cite{guglietta2020effects,guglietta2024analytical}, enabling a more complete assessment of the EMM model across its full parameters space.
Fig.~\ref{fig:error} reports $\averr$ as a function of $\Bqs$ and $\Bqd$, for different choices of $\Ca$. As a preliminary remark, we note that the error maps are generally not symmetric with respect to the diagonal $\Bqs=\Bqd$: exchanging $\Bqs$ and $\Bqd$ does not yield the same error. This asymmetry is consistent with the analytical structure of the coefficients $f_{1,2}^{(\mbox{\tiny EMM})}$ (see Eq.~\eqref{ourf1f2} and Fig.~\ref{fig:f1}).
The resulting colormaps reveal that, in general, at small interfacial viscosities ($\Bqs,\Bqd \ll 1$), the EMM model reproduces FRS data with excellent fidelity, with errors typically below a few percent. For relatively small capillary numbers ($\Ca\le0.2$), as either $\Bqs$ or $\Bqd$ grows, the error increases: however, the degradation is modest along the diagonal $\Bqs=\Bqd$ for intermediate values of Boussinesq numbers, consistently with the excellent agreement observed in Fig.~\ref{fig:D_vs_t}, panel (b).  
In contrast, strongly asymmetric viscosity configurations (where either $\Bqs$ or $\Bqd$ dominates), lead to noticeably larger discrepancies.
Increasing the capillary number $\Ca$ pushes the model closer to its limits. At $\Ca=0.3$ and 0.5, a sharper growth of the error appears, especially when both viscosities are large. This behaviour is expected: large deformations amplify nonlinear couplings in the interfacial stress that the EMM model, derived in the small-deformation limit, can only approximate. {Overall, 
Fig.~\ref{fig:error} provides a quantitative map of the domain of validity of the EMM model, highlighting the regions where deviations grow and FRS become indispensable. 
Note that, at $\Ca=0.5$, breakup occurs for small values of $\Bqs$ (the corresponding region is shown in grey). {Although the FRS does not resolve topological changes, the onset of breakup is still captured in practice as a loss of numerical stability at large deformations~\cite{guglietta2025Deformation}.} This behaviour is fully consistent with the physical mechanism discussed above: when $\Bqs=0$, the surface-divergence term in the interfacial stress remains active and enhances the deformation, making the droplet more susceptible to breakup. As $\Bqs$ increases, this contribution is progressively damped, leading to a stabilizing effect.

%%%%%%%%%%%%%%%%%%%%%%%%%%%%%%%%%%%%%%%%%%%
\section{Conclusions and perspectives}\label{sec:conclusions}
%%%%%%%%%%%%%%%%%%%%%%%%%%%%%%%%%%%%%%%%%%%%
{We have studied a reduced model to investigate the dynamics of a droplet with interfacial viscosity in shear flow. The model extends the Maffettone–Minale (MM) framework~\cite{maffettone1998equation} (Eq.\eqref{eq:MMmodel}) by supplementing surface-tension effects with interfacial viscosity, under the assumption of Newtonian interfaces~\cite{Scriven1960,langevin2014rheology}. The resulting extended MM (EMM) model provides a simple phenomenological description of time-dependent droplet deformation while retaining a clear connection with classical small-deformation theories~\cite{frankel1970constitutive,flumerfelt1980effects,barthesbiesel1985,narsimhan2019shape} and allowing an independent control of shear and dilatational interfacial viscosities through the corresponding Boussinesq numbers $\Bqs$ and $\Bqd$.}

A systematic comparison with fully resolved simulations (FRS), performed using an immersed boundary–lattice Boltzmann approach~\cite{art:lizhang19,guglietta2020effects,guglietta2025Deformation}, delineates the regimes in which the EMM model remains quantitatively accurate. The comparison highlights how the stationary deformation $D_S$ and the range of capillary numbers $\Ca$ over which the linear response theory applies depend sensitively on the interfacial viscosities. While for strongly asymmetric viscosity configurations the linear relation between $D_S$ and $\Ca$ persists over a relatively broad interval of $\Ca$, this range is significantly reduced in the equiviscous case $\Bqs=\Bqd$, where nonlinear effects become relevant already at moderate deformations. The analysis of the time-dependent deformation $D(t)$ further shows that interfacial viscosity strongly affects the transient dynamics: shear and dilatational viscosities slow down the deformation process in a nonlinear and coupled manner, with the most pronounced effect occurring when the two viscosities are comparable. This behaviour is quantitatively captured by the characteristic deformation time, $\tau_\sigma/f_1^{(\mbox{\tiny EMM})}$, which is directly controlled by the coefficient $f_1^{(\mbox{\tiny EMM})}$, and increases sharply only when both $\Bqs$ and $\Bqd$ are appreciable. The error maps in Fig.~\ref{fig:error} further provide a global view of the model’s domain of validity in the $(\Ca,\Bqs,\Bqd)$ space. For moderate deformations (small to intermediate $\Ca$) and in the equiviscous case $\Bqs=\Bqd$, the EMM model reproduces FRS results with errors of only a few percent. Larger discrepancies arise for strongly asymmetric interfacial viscosities or at larger $\Ca$, consistent with the small-deformation assumptions underlying the model and the absence of higher-order nonlinear interfacial effects. {Overall, these results show that the EMM model constitutes a robust and computationally inexpensive tool for predicting droplet deformation with interfacial viscosity.}

While the present study focuses on a simple shear flow, the EMM framework can be naturally extended to other flow configurations and to regimes of larger deformation, including the prediction of the critical capillary number for breakup, $\Ca_{\mathrm{cr}}$~\cite{maffettone1998equation}. Although quantitative accuracy cannot be expected at large $\Ca$, the availability of analytical expressions for $\Ca_{\mathrm{cr}}$ in terms of $f_{1,2}$ makes it possible to assess deviations from established results~\cite{art:gounley16,singh2020deformation}. {Following Ref.~\cite{taglienti2023reduced}, one could also determine modified coefficients $f_{1,2}$, calibrated for finite-$\Ca$ conditions.}
Another promising direction concerns the link between interfacial viscosity and surfactant-laden interfaces. Surface viscous stresses are often associated with the presence of surfactant monolayers~\cite{PonceTorres2017surfact,luo2019influence,wee2020effects,herrada2021stability}, suggesting that the EMM framework could be used to extend earlier studies~\cite{biferale2014deformation} and infer droplet deformation statistics in complex flows {(as in Ref.~\cite{Taglienti2024,guglietta2025Deformation})} with surfactants. This would require coupling the flow dynamics to surfactant transport and to an equation of state relating surface concentration to interfacial viscosities~\cite{PonceTorres2017surfact,manikantan2017pressure}. Since surfactants also exhibit intrinsic interfacial dynamics~\cite{brooks1999interfacial,verwijlen2011study,fuller2012complex,samaniuk2014micro,zell2014surface,zell2016linear}, an open question is to what extent the MM framework can be further generalized to incorporate such surface heterogeneities while retaining a reduced description.

\acknowledgments
We acknowledge Fabio Bonaccorso for support. This work received funding from the European Research Council (ERC) under the European Union's Horizon 2020 research and innovation programme (grant agreement No 882340). This work was supported by the Italian Ministry of University and Research (MUR) under the FARE program (No. R2045J8XAW), project ``Smart-HEART''. MS acknowledges the support of the National Center for HPC, Big Data and Quantum Computing,  Project CN\_00000013 - CUP E83C22003230001,   Mission 4 Component 2 Investment 1.4, funded by the European Union - NextGenerationEU. Support/funding from Tor Vergata University project AI4HEART and INFN/FIELDTURB project are also acknowledged. 
\bibliography{biblio}

\end{document}